\documentclass[english,pra,amsfonts,amssymb,amsmath,twocolumn,groupedaddress]{revtex4-1}
\usepackage[utf8]{inputenc}
\pdfpageattr{/Group <</S /Transparency /I true /CS /DeviceRGB>>}
\usepackage{amsthm}
\usepackage{amsmath}
\usepackage{amssymb}
\usepackage{subcaption}

\PassOptionsToPackage{normalem}{ulem}
\usepackage{ulem}

\usepackage{comment}
\usepackage{caption}
\usepackage{algpseudocode}

\algnewcommand{\A}{\textbf{and}\space}
\algnewcommand{\Or}{\textbf{or}\space}
\algnewcommand{\Xor}{\textbf{xor}\space}
\makeatletter
\let\OldStatex\Statex
\renewcommand{\Statex}[1][3]{%
  \setlength\@tempdima{\algorithmicindent}%
  \OldStatex\hskip\dimexpr#1\@tempdima\relax}
\makeatother

\usepackage[unicode=true,pdfusetitle,
 bookmarks=true,bookmarksnumbered=false,bookmarksopen=false,
 breaklinks=false,pdfborder={0 0 1},backref=false,colorlinks=false]
 {hyperref}
\usepackage[usenames,dvipsnames]{xcolor}
\hypersetup{colorlinks=true, linkcolor=Maroon, citecolor=OliveGreen, filecolor=magenta, urlcolor=Blue}

\makeatletter

\theoremstyle{plain}

\theoremstyle{plain}

\makeatother

\usepackage{babel}

\providecommand{\theoremname}{Theorem}
\providecommand{\definname}{Definition}
\providecommand{\observname}{Observation}
\providecommand{\corolname}{Corollary}
\providecommand{\algorithmname}{Algorithm}
\providecommand{\examplename}{Example}
\providecommand{\problemname}{Problem}

\usepackage{pgfplots}
\usetikzlibrary{calc}
\usepackage{pgfplotstable}
\usepackage{colortbl}
\usepackage{booktabs}
\usepackage{pgfplotstable}

\begin{document}

\title{Predicting human-generated bitstreams using classical and quantum models}

\author{Alex Bocharov$^1$}
\author{Michael Freedman$^{2,3}$}
\author{Eshan Kemp$^1$}
\author{Martin Roetteler$^1$}
\author{Krysta M.~Svore$^1$}

\affiliation{$^1$ Microsoft Quantum, Redmond, WA (USA)}
\affiliation{$^2$ Station Q, Microsoft Research, Santa Barbara, CA (USA)}
\affiliation{$^3$ Department of Mathematics, University of California, Santa Barbara, CA (USA)}

\date{\today}

\keywords{Quantum Computing, Machine Learning, Quantum Circuits}

\vskip 0.3in

\begin{abstract}
A school of thought contends that human decision making exhibits quantum-like logic. While it is not known whether the brain may indeed be driven by actual quantum mechanisms, some researchers suggest that the decision logic is phenomenologically non-classical. This paper  develops and implements an empirical framework to explore this view. We emulate binary decision-making using low width, low depth, parameterized quantum circuits. Here, entanglement serves as a resource for pattern analysis in the context of a simple bit-prediction game. We evaluate a hybrid quantum-assisted machine learning strategy where quantum processing is used to detect correlations in the bitstreams while parameter updates and class inference are performed by classical post-processing of measurement results. Simulation results indicate that a family of two-qubit variational circuits is sufficient to achieve the same bit-prediction accuracy as the best traditional classical solution such as neural nets or logistic autoregression. Thus, short of establishing a provable ``quantum advantage'' in this simple scenario, we give evidence that the classical predictability analysis of a human-generated bitstream can be achieved by small quantum models.
\end{abstract}

\maketitle

\section{Introduction}
\label{sec:intro}

There has been a scholarly discussion, going back at least to letters exchanged by Wolfgang Pauli and Carl Jung in the 1930s, on the relation between the mind and the quantum world. This question has also been the subject of provocative, if not wild, hypotheses: Roger Penrose famously proposed our brains employ quantum gravity. Although no fully satisfactory physical linkage between the known classical appurtenances of the brain with a hypothetical quantum layer have been found, scientific work on the topic advances~\cite{fisher15}.

A line of evidence is drawn from certain psychological paradoxes (e.g., the ``Ellsberg Paradox'') where subjects eschew classical logical concepts, as evidenced through their decisions, but instead make choices that can be modeled with the help of ``non-commuting operators'', a staple of the quantum world (cf.~\cite{aerts15}, \cite{halpern17}).

Our approach is to be agnostic regarding the ambitious question of ``Does quantum information play a role in brain function?'' Instead, we aim at providing evidence that it is possible to train quantum mechanical models that have predictive power in the realm of human decision making.

To this end, we consider a limited model of decision-making in which a human plays a simple game against a computer that tries to predict the human's next move. The game is a binary version of ``rock, paper, scissors,'' consists of $n$ rounds, where $n$ is large enough to allow meaningful prediction of patterns. In each round the computer makes a binary decision with the outcomes labeled $0$ and $1$ and stores the value $c$ of this bit. After that a human player makes the same kind of decision, stores the value $b$ of their bit, which is then compared to the computer's choice. Computer wins if and only if $c\oplus b = 0$, or in other words, if the computers decision correctly anticipates the human's decision.

\begin{figure}[h!]
\includegraphics[width=3.2in] {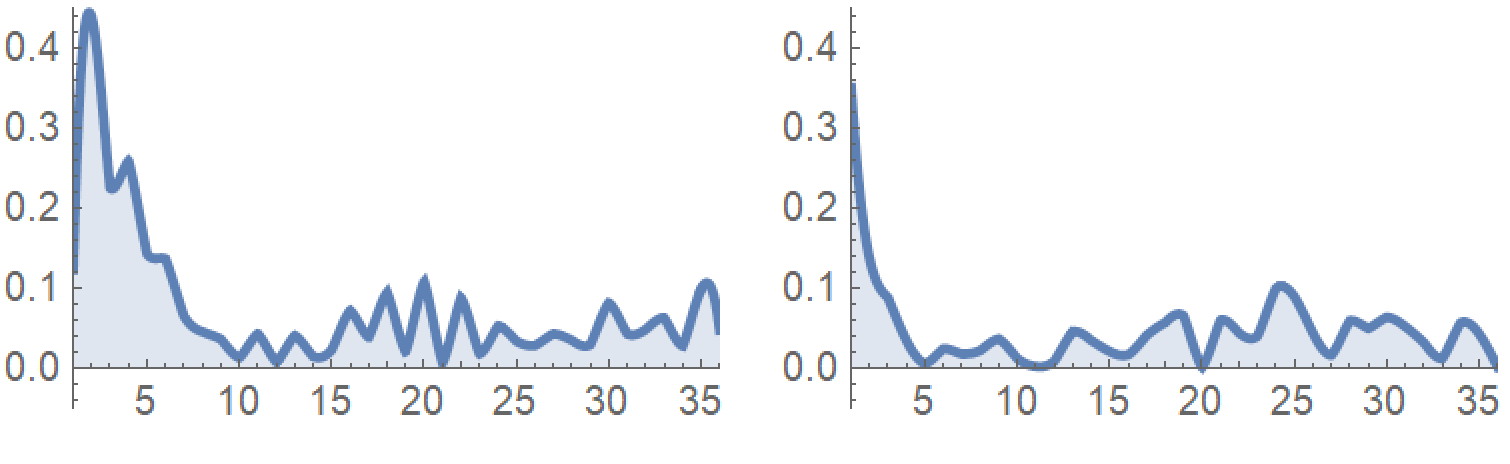}
\caption{\label{fig:autocor} Typical behavior of autocorrelation strength $\{a_s\}_{s=0}^{n-1}$ for two binary sequences $\{b_t\}_{t=0}^{n-1}$, shown for two sample sequences of length $n=1000$ that were entered by humans. Here $a_s$ is defined as absolute value of Pearson correlation coefficient $a_s = |\rho(b_t,b_{t-s})|=|\mbox{Cov}(b_t,b_{t-s})/(\sigma(b_t)\,\sigma(b_{t-s}))|$. Random sequences satisfy $\mathbf{E}[a_s] = \frac{1}{\sqrt{s}}$, which for the above length would be around $0.032$. The quickly decaying autocorrelations (only the first $36$ taps are shown) suggest that small scale correlations dominate in sequences generated by humans who try to behave as random as possible. This opens up the possibility to forecast the next element in a bitstream sequence with a probability that significantly exceeds 0.5.}
\end{figure}

It is assumed that the human player does not has access to any mechanical or electronic random number generators and thus have to rely solely on their minds to make the binary decisions. The computer is not constrained on the amount of randomness it can use as a resource. Clearly, having access to an unbiased coin that generates a uniform distribution $p(b=0)=p(b=1)=\frac{1}{2}$ allows to win this game an expected number of $n/2$ rounds, and this is true against any strategy. However, if a sequence of bits (i.e., a bitstream) is generated by a human, the bits typically are far from being independent, identically distributed, and unbiased random variables. Fig.~\ref{fig:autocor} shows the autocorrelation function of a few samples of sequences of length $1000$ that were entered by a group of volunteers for this study.

Classical approaches to bit-prediction have a long history, see e.g.~\cite{merril2018} for an implemention based on $d$-grams that was conceived by Scott Aaronson. In order to explore quantum-assisted alternatives of bit-prediction we experimented with a hybrid quantum-classical approach based on the extensions of quantum classifier circuits proposed in \cite{schuld18}.
A quantum classifier circuit is a parameterized rapidly entangling circuit that is using a quantum state encoding of a classical data vector and is striving to make a $0/1$ decision on said data by measuring a certain observable with eigenvalues $\pm 1$. The parameters of this circuit are learned from the snippets of human-generated bitstreams using stochastic gradient descent \cite{bottou2004} or more robust training alternatives.
We considered different encodings of the bitstreams as quantum states, such as qubit encoding and amplitude encoding \cite{schuld18}, and combined these with different training methods, such as stochastic gradient descent and coordinate ascent training. Amplitude encoding resulted in a rather simple two-qubit quantum circuit with just eight trainable parameters that performs on a par with a suite of classical solutions that we compare our method with.

\section{Hybrid Quantum-Classical Approach}
\label{sec:hybrid}

\subsection{Predictor Design} \label{subsec:predictor}

We define forecasting the next human's choice at time $t$ given the history of their previous choices as the task of sampling from the conditional probability distribution
\begin{equation}
P_t = p(b_t=1 | b_{t-1}, b_{t-2},\ldots,b_0),
\end{equation}
\noindent where $b_{t-s}$ is the chosen bit at the round $t-s$.

We assume that the correlation between $b_t$ and $b_{t-s}$ decays exponentially as $s$ grows indefinitely and therefore, for practical purposes, there exists some effective depth $d$ such that
$p(b_t=1 | b_{t-1}, b_{t-2},\ldots,b_{t-d})$ is a good approximation for the $P_t$ for large enough values of $t$.

Following the recipes proposed in \cite{schuld18} it makes sense to explore two possible encoding methods for the bitstream short memory $m_{t,d}=[b_{t-1}, b_{t-2},\ldots,b_{t-d}]$. The first method uses $k=d$ qubits and encodes the $m_{t,d}$ as the pure state $\psi_t = |b_{t-1}, b_{t-2},\ldots,b_{t-d}\rangle$ in standard computational basis; the other method uses $k=\lceil \log_2(d)+1 \rceil$ qubits and employs \emph{amplitude encoding}
\[
\psi_t = \nu \left( |0\rangle + \sum_{s=1}^d b_{t-s} |s\rangle \right)
\]
\noindent where $\nu$ is normalization factor so that $\| \psi_t \|_2 = 1$.

We then interpret $P(b_t=1 | b_{t-1}, b_{t-2},\ldots,b_{t-d})$ as the probability of measuring eigenvalue $-1$ of a fixed  parameterized observable on the $k$-qubit  register.

More precisely, we take an equivalent view of the measurement and interpret $p(b_t=1 | b_{t-1}, b_{t-2},\ldots,b_{t-d})$ as the probability of measuring $-1$ (in the standard basis) on one of the qubits in the state $U(\theta_t) \psi_t$,
where $U(\theta_t)$ is a parameterized unitary on the $k$-qubit register with polynomially many learnable parameters $\theta_t$.

In this model we interpret the learning of the human behavior as learning of the parameters of the $U(\theta_t)$ transform. For the learning goal: let us view $\psi_t$ as a data case and the bit $b_t$ as its label. Let us interpret sampling for a forecasted bit $\hat{b}_t$ as sampling for the class label. This maps the bit forecast task onto a classification task and learning of $U(\theta_t)$  into a supervised learning of binary classifiers.

The utility function for both tasks is the same:
\begin{equation} \label{eq:utility:function}
L(\theta_t) = \sum_t \langle U(\theta_t)  \psi_t | \Pi_{b_t} |  U(\theta_t)  \psi_t \rangle
\end{equation}
\noindent where $\Pi_b$ projects on the $(-1)^{b}$ eigenspace of $Z\otimes I^{k-1}$.

Since $\Pi_b = 1/2 (I^{k} + (-1)^b Z\otimes I^{k-1}); b=0,1$, then
\begin{equation} \label{eq:utility:rewritten}
L(\theta_t) = \mbox{c} + \frac{1}{2} \sum_t (-1)^{b_t} \langle U(\theta_t)  \psi_t |Z\otimes I^{k-1} |  U(\theta_t)  \psi_t \rangle.
\end{equation}
In practice, learning parameters $\theta_t$ as an optimal point of $L(\theta_t)$ is often done by \emph{stochastic gradient descent} strategy.
In order to get less chaotic and predictable gradient updates, it is a common practice to create ``mini batches'' of consecutive terms.

That is for some small mini batch count $\rm{mb}$ we use the following parameter update rule that replaces $\theta_t$ with
\[
\theta_t + \lambda \, \sum_{\tau = t-\rm{mb}}^{t} \left( (-1)^{b_{\tau}} \, \nabla_{\theta_t} \langle U(\theta_t)  \psi_{\tau} | Z\otimes I^{d-1} |  U(\theta_t) \psi_{\tau} \rangle \right)
\]
\noindent where $\nabla_{\theta_t}$ is the gradient and  $\lambda$ is the \emph{learning rate}.

As we have discovered empirically, using stochastic gradient descent in this context is costly and inconvenient. We have instead used a more recent strategy for optimization of variational quantum circuits known as \emph{coordinate ascent}, cf.  \cite{OstaszewskiEtAl} and \cite{toappear20}.

In short, the coordinate ascent method is applicable to circuits that are composed of \emph{Pauli rotations} $exp(-i\,\theta\, P)$,  where $P$ is some Pauli operator, $P^2 = I^k$,  and generalized controlled Pauli rotations $\Pi^{\perp}+\Pi \, exp(-i\,\theta\, P)\, \Pi$ there $P$ is a Pauli operator and $(\Pi,\Pi^{\perp})$ is a pair of complementary orthogonal projectors with $\Pi^2=\Pi, \Pi^{\perp}=I^k-\Pi$.
In particular the \emph{polar code} circuits described below are explicitly seen as compositions of such gates.

The premise in the coordinate ascent strategy is that if the values of all but one the circuit parameters $(\theta_1,\ldots,\theta_{j-1},\theta_{j+1},\ldots,\theta_L), j\in [L]$ are considered fixed then the conditional absolute (arg)maximum
of a likelihood function such as (\ref{eq:utility:rewritten}) in the single variable $\theta_j$ is obtained in closed form at constant cost.

\begin{figure}
\includegraphics[width=3.3in] {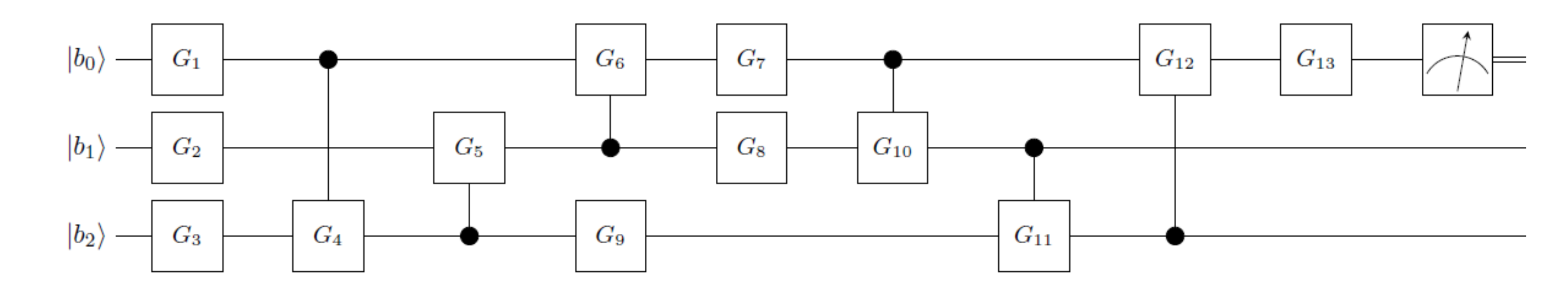}
\caption{\label{fig:polar:code} Rapidly-entangling 3-qubit circuit of depth 9 and size 13. Each of the gates $G_j$, where $j=1,{\ldots},13$, is a single qubit quantum gate specified by 3 real parameters. In practice, restricting to learnable rotations with just 2 real parameters appears to provide sufficient expressive power for the classifier to train and generalize well.}
\end{figure}

In order to make the training process classically amenable we employ a specific parsimonious representation for the unitary transform $U(\theta)$ in the form of \emph{polar code} circuit. An example of such circuit for $k=3$ is shown on the FIG \ref{fig:polar:code}. All the gates $G_j$, where $j \in \{1,\ldots,13\}$ are single-qubit gates. The controlled gates $G_k$, where $k \in \{4,\ldots,6\} \cup \{10,\ldots,12\}$ are set up to provide near-maximal entanglement/unentanglement capacity which allows to represent the intra-data correlations at various ranges. (See \cite{fourAuthors19}, \cite{deng17} for insights on the entanglement as a resource for representing correlations.)

For a given set of parameters and the most recent bit $d$-gram $B_t = [b_{t-1}, b_{t-2},\ldots,b_{t-d}]$ the bit forecasting circuit must be set up and run several times in order to ensure bit forecast based on representative sample of the conditional distribution $p(b_t=1 | b_{t-1}, b_{t-2},\ldots,b_{t-d})$. In \cite{schuld18} section IV.E.2 we described the sample size needed for estimating that conditional distribution to a given precision. A more simplistic approach is to run the circuit $S$ times all the way through the measurement and then select the forecasted bit by majority of measurement results. However, even with this simplification we need some bounds (especially, the lower bound) on $S$ in order to ensure the robustness of the majority vote.

Suppose, as above, that $\psi_t$ is a chosen quantum encoding of the $d$-gram $B_t$. For the estimate of the conditional probability on a quantum device one can use the \emph{Hadamard test} which computes the overlap
$\langle \psi_t| U(\theta_t)^{\dagger} (Z\otimes I^{k-1}) U(\theta_t) | \psi_t \rangle$ by preparing additional non-informative ancillary qubit, indexed with $0$ w.l.o.g., in the state $|+\rangle$ running the controlled version $C(U(\theta_t)^{\dagger} (Z\otimes I^{k-1}) U(\theta_t))$ on  $|+\rangle |\psi_t\rangle$ and comparing the probabilities of measuring $1$ and $-1$ on the ancilla in the resulting state. Hadamard test is described in \cite{AJL2005}.
Suppose at the point of time $t$ the next bit $b_t$ is highly forecastable, which would mean, for example that $p_1 = p(b_t=1 | b_{t-1}, b_{t-2},\ldots,b_{t-d}) \gg p(b_t=0 | b_{t-1}, b_{t-2},\ldots,b_{t-d}) = p_0$. Let us set $\varepsilon = (p_1-p_0)/2$. Suppose $S$ is the number of samples from the distribution sufficient for estimating $p_0$ and $p_1$ to precision $\varepsilon$; then $S$ is the number of reruns of the classifier circuit sufficient for a robust majority vote. Indeed, given the above description, it is highly unlikely that $0$'s are going to be in the majority among the $S$ samples. Here for completeness we give a short description of a method for estimating the stochastic gradient, while referring to \cite{schuld18} for details.

An approximation for the gradient $\nabla_{\theta_t} \langle U(\theta_t)  \psi_t | Z\otimes I^{k-1} |  U(\theta_t) \psi_t \rangle$ can be obtained using overlap estimators for a set of coherent unitary circuits, closely related to $U(\theta_t)$. To this end, suppose for simplicity that
\begin{equation}
U(\theta_t) = U_1(\eta_1) \cdots U_L(\eta_L)
\end{equation}
\noindent where each $U_{\ell}(\eta_{\ell})$ is a unitary depending on only one subparameter $\eta_{\ell}$ and $\eta_1, \ldots, \eta_L$ are all distinct.

By direct computation,
\begin{eqnarray*} \label{eq:partial:eta}
\partial_{\eta_j}(\langle U(\theta_t)  \psi_t | Z\otimes I^{k-1} |  U(\theta_t)  \psi_t \rangle) = \\
2 \Re(\langle U_1 \cdots \frac{\partial U_j(\eta_j)}{\partial \eta_j} \cdots U_L  \psi_t | Z\otimes I^{k-1} |  U(\theta_t)  \psi_t \rangle).
\end{eqnarray*}

We further note that, whenever $U_j(\eta_j)$ is an axial single-qubit rotation by the angle $\eta_j$, then $\frac{\partial U_j(\eta_j)}{\partial \eta_j}$ is also a rotation about the same axis by a deterministically modified angle. Therefore the right hand side is obtained as an overlap of two unitary states across one projector to an eigenspace of $Z\otimes I^{k-1}$.

If $U_j(\eta_j)$ is a \emph{controlled} single-qubit rotation, then $\frac{\partial U_j(\eta_j)}{\partial \eta_j}$ is not a unitary gate. However, it is a \emph{linear combination} of two unitary gates:
\[
\frac{\partial U_j(\eta_j)}{\partial \eta_j} = \frac{1}{2}(I \otimes \frac{\partial V_j}{\partial \eta_j} - Z \otimes \frac{\partial V_j}{\partial \eta_j} )
\]
\noindent where each of the two terms on the right can be treated by running a purely unitary circuit.

\subsubsection{Gradient-free coordinate ascent}
A more robust alternative to the gradient descent is a strategy of sequential likelihood maximization, where for a selected parameter index $j \in [L]$ we deem all the parameters, except $\eta_j$ fixed and we use explicit equations for $\exp(-i \, \eta_j \, P_j)$ to obtain conditional absolute maximum of the likelihood in $\eta_j$ in closed form, see~\cite{toappear20}. As a result, coefficients appearing in the equations for the conditional $\mbox{argmax}$ in $\eta_j$ are all \emph{quantum overlaps} of the form either
\[
\Re \langle V_{j,t} \psi_t | O_{j,t} | W_{j,t} \psi_t \rangle \, \mbox{or} \, \Im \langle V_{j,t} \psi_t | O_{j,t} | W_{j,t} \psi_t \rangle
\]
\noindent where $V_{j,t},W_{j,t}$ are certain sub-circuits of the circuit $U(\theta_t)$ and $O_{j,t}$ is a simple explicit observable, usually just a Pauli $Z$ on one of the qubits.

Both the real and the imaginary parts of a quantum state overlap can be estimated using two complementary versions of the Hadamard test.

\subsubsection{Multi-epoch training}

Due to randomized nature and relatively slow convergence of the stochastic gradient descent strategy, the usual practice in stochastic learning is to make multiple passes through the training data, which means in our case through a significant segment of the bit history $b_{t-1}, b_{t-2},\ldots,b_0$.

When using coordinate ascent as an alternative to gradient descent, obviously, we need to touch all or most of the circuit parameters at least once, thus making $\Omega(L)$ optimization steps.
Empirically it is evident that the number of passes scales as $O(\log(1/\epsilon))$ where $\epsilon$ is the desired precision. The coordinate ascent strategy appears to be more robust compared to the gradient descent, since the number of epochs required for the convergence of the latter strongly depends on hyperparameters and is hard to predict.

\subsubsection{Random restarts}
The landscape of the goal function (\ref{eq:utility:rewritten}) over the parameter space is pronouncedly non-convex. In order to increase the chances of finding a good local optimum, multiple initializations of the starting parameter vector have been considered and evaluated in parallel.
goal function was then selected for validation and evaluation.

\subsection{Data and methodolgy}

We obtained an experimental proof of concept for the solution described in the previous subsection by coding all the circuits involved in Q\# and running them on the Microsoft Quantum Developmenet Kit~\cite{QSharp2019}.

\subsubsection{Test data: synthetic and human-entered}

Practical experimentation was performed on both synthetic and humanly-generated data. Synthetic data was generated using some deterministic rules and a certain levels of randomization.

In most of the generality the synthetic data generator can be described as a randomized regression $b_t = a_1\, b_{t-1} + \cdots + a_k \, b_{t-k} + r$, where $a_j \in \{0,1\}, j = 1..k$ and $r$ is a random bit drawn from some skewed distribution. All additions are modulo 2. Therefore, there is a deterministic bit $a_1\, b_{t-1} + \cdots + a_k \, b_{t-k}$ depending on the bitstream history of depth $k$ that can be flipped with a certain probability $P(r=1)$.
The actual sequence generated is defined by the regression equation and the $k$ initial seed bits $b_{k-1},\ldots, b_0$.
It is known that, in absence of the noise bit $r$ the above regression
 generates a periodic sequence with the period of at most $2^k$. Synthetic data was used for training tune-ups.

In addition to the synthetic data, we targeted two different settings for collecting human-generated data. For the first setting,
 we created an interactive application, where either a classical $d$-gram oracle or the simulated hybrid quantum predictor (as described in subsection \ref{subsec:predictor}) was randomly selected to play against the human. In order to bring in some psychology the intra-round gains/losses were measured in dollars. There was a certain maximal gain titled ``jackpot'' and a certain maximal loss titled as ``being broke''. We collected over one hundred bitstreams from volunteers playing against this application. We refer to these bitstreams as \emph{game transcripts}.

In the second setting, each of our $34$ volunteers was asked to produce a string of 1000 bits while keeping it ``as random as possible''. Volunteers produced 32 bit strings of this kind in a single data collection session. We refer to these data samples as \emph{simple bitstreams}.
		
\subsubsection{Qualitative observations on the data}

Even though the game transcript data had been collected interactively we disregarded its interactive genesis in these bitstreams investigation and focused on post-mortem analysis of their predictability.
Psychologists (cf.~\cite{rng2008}, \cite{rng2012}) were noting earlier that in absence of mechanical aid an average human is not too good at maintaining fair randomness. It appears that, in time a human subject tends to form  a subconscious pattern that biases his or her choices.
Contrary to an anecdotal claim in \cite{merril2018}, we determined, however (see results in the subsequent sections of this paper) that the achievable average accuracy is closer to 64 percent, at least in the setting, where the subjects were instructed to ``randomize''. We also witnessed a handful of subjects who managed to achieve a near-perfect degree of randomization: on their bitstreams no predictor was performing better than a fair coin toss.

Somewhat surprisingly, we are seeing little difference between statistical properties of the ``game transcripts'' and those of the ``simple bitstreams''. It appears that informing subjects with the running gain/loss feedback against computed predictions does not have certifiable impact on their ability to randomize. (We have seen signs that human behavior becomes somewhat more predictable close to ``being broke'' cutoff, but could not establish this with sufficient statistical significance.)

\subsubsection{Accuracy evaluation methodology} \label{subsub:evaluation:methodology}
In order to collect unbiased and comprehensive statistics, a subsequence of $L$ consecutive bits (``training window'') was extracted from each test bitstream for the purposes of model training; then the $5$ bits immediately following the training window were used for the accuracy scoring.

The training windows were \emph{staggered} across the test population. That is a training window in a test bitstream  $b_0, \ldots, b_T$ was selected as $b_{\tau+1},\ldots,b_{\tau+L}$, where the offset $\tau$ was drawn from a uniform distribution over $[0,T-L-5]$. Thus our scoring approach called for maximizing the probability of computer win ``anytime anywhere in the game''. In this scheme, a population of $M$ test bitstreams would yield a total of $5\,M$ win/loss bits. The predictor accuracy score was then estimated as $(\#\mbox{wins})/(5\,M)$.

\subsection{Quantum and classical benchmarks}

User-generated \emph{game transcripts} (i.e. bitstreams collected interactively) turned out, a posteriori, to be statistically similar to the ``simple bitstreams'' (collected without computer interaction).

We used overall training window width $L$, as described in subsection \ref{subsub:evaluation:methodology}, as one of the key benchmarking hyperparameters. After several rounds of experimentation we observed that most competing predictor designs, both traditional classical and circuit-centric quantum, perform significantly better for $100 \leq L \leq 150$. (The lower bound $100 \leq L$ availed the corresponding predictors enough training data, whereas the $L \leq 150$ was likely the statistical stationarity horizon in a typical bitstream segment.) Accordingly at the second stage of experimentation shorter game transcripts have been removed from consideration and only bitstreams with $125$ or more have been retained.

\paragraph{Observation.} Most of the donated bitstreams tend to show small individual bias towards entering $0$ bit. Denoting the frequency of $0$ bit in a stream $s$ by $f_0(s)$ we find that $f_0$ is distributed across the streams as roughly $N(0.515,0.03)$. Thus, the majority of the donated bitstreams turned out to be asymmetric in this respect.

\subsubsection{Conditional collision statistics} \label{sub:sub:collision:statistics}

Recall that given a binomial distribution $(p_0,p_1),p_0+p_1=1$, the quantity $p_0^2+p_1^2$ is called the \emph{collision probability} of the distribution. Accordingly, given a selected bitstream depth $d$ and considering the conditional distribution $(p_0(t,d)=p(b_t=0|b_{t-1},\ldots,b_{t-d}), p_1(t,d)=p(b_t=1|b_{t-1},\ldots,b_{t-d}))$, we call
$cp(t,d)=p_0(t,d)^2+p_1(t,d)^2$ the \emph{conditional collision probability} of the stream $d$-grams at time point $t$.
By and large the conditional collision probabilities are not directly observable and must be estimated. For the particular data and a training window $W(L)$ of width $L$ we introduce conditional counts
$C_j(b_{t-1},\ldots,b_{t-d},W(L))=\#occur([j,b_{t-1},\ldots,b_{t-d}] \subset W(L)), j=0,1$. For a $d$-gram $\bar{b}$ we next introduce
\begin{eqnarray*} \label{eq:conditional:frequencies}
\hat{p_j}(\bar{b},W(L)) =C_j(\bar{b},W(L))/(C_0(\bar{b},W(L))+\\
C_1(\bar{b},W(L))), j = 0,1
\end{eqnarray*}
\noindent and the \emph{conditional collision frequency}
\[
\hat{cp}(\bar{b},W(L)) = \hat{p_0}(\bar{b},W(L))^2 + \hat{p_1}(\bar{b},W(L))^2.
\]
Finally, we propose here a model-free inference strategy for inference of the follow-on bits given a $d$-gram $\bar{b}$: we sample the inferred bit randomly from the binomial distribution
$(\hat{p_0}(\bar{b},W(L)),\hat{p_1}(\bar{b},W(L)))$. (To the best of our knowledge, such inference strategy is used by the $d$-gram oracle \cite{merril2018}, except that the latter does not have a constraint
on the window width $L$.)

It is easy to see that the conditional collision frequency $\hat{cp}(\bar{b},W(L))$ is an unbiased estimate for the expected probability for inferring the follow-on bit for $\bar{b}$ \emph{correctly} using the above strategy. In that sense the $\hat{cp}((\bar{b},W(L))$ benchmarks the expected accuracy of model-free inference strategies.

\paragraph{Observation.}
For the human-generated bitstream data set and $100 \leq L \leq 150$ the $\hat{cp}(\bar{b},W(L))$ was distributed across the set of all $(\bar{b},W)$ as $N(cp,\sigma)$ where $cp$ was in the ballpark of $0.6$ and $\sigma$ did not exceed $0.08$. In particular for $L=125$ we estimated $cp=0.62$ and $\sigma=0.07$ with statistical significance $0.98$.

This is an early evidence that the binary choices had been not completely random and had been somewhat predictable in the majority of cases. It also sets a bar for required accuracy of specific predictive models in this context. It turns out that fashioning a predictive model that exceeds the above mentioned 62 percent average is not trivial.

\subsubsection{Classical predictors}

In order to create representative classical benchmarks for evaluation of proposed quantum designs, we explored a collection of publicly and commercially available predictive packages.

 In addition to the $d$-gram add hoc implementation of $d$-gram model free inference, as described in subsection \ref{sub:sub:collision:statistics},  we selected several feed-forward neural net (FFNN) classifiers supported by the Python $\mbox{scikit-learn}$ package. For this purpose the $\mbox{scikit-learn}$ package implements the $\mbox{MLPClassifier}$ class \cite{scikit2019}
 with selectable hidden layer sizes. We limited our choices to geometries with at most 3 hidden layers as MLPs with more hidden layers tend to overfit and undergeneralize.
 To build a solution for $d$-gram depth $d$ we evaluated the following $6$ choices of the {\tt hidden\_layer\_sizes} ($HLS$) parameter for the subject MLPClassifier instance: $HLS \in \{(d),(d,d),(d,d,d),(f),(f,f),(f,f,f)\}$, where $f=\lfloor 2\,d/3 \rfloor$, which is in line with commonly adopted neural net heuristics.

 We performed data analysis sweeps using there types of activation options: ReLu, Softmax and Tanh. We relied on default regularization settings.

 At the second stage of experimentation we included commercial machine learning packages released by Wolfram \emph{Mathematica} edition 12 that offers the $\mbox{Classify[]}$ function with a broad choice of predictive engines \cite{mmaClassify2019} such as ``LogisticRegression'', ``NaiveBayes'', ``NearestNeighbors'',``RandomForest'', ``SupportVectorMachine''. We performed full sweep across all these choices.

\subsubsection{Quantum predictors}

We employed the circuit-centric quantum  (QCC) predictors in our simulations. The first distinction as defined in the predictor design section \ref{subsec:predictor} was between the \emph{qubit encoding} and \emph{amplitude encoding} of the bit string short memories (the $d$-grams). The major top level distinction between the first and the second stage of experimentation was the use of the stochastic gradient descent
versus coordinate ascent training.

On all the stages we used the same quantum circuit geometry similar to one shown on FIG.\ref{fig:polar:code}.
The simulations had been performed across a matrix of varying circuit widths and depths.
The variation of depth of the quantum circuit of the QCC model was achieved by replicating entangling blocks.
Given a quantum register with $k$ qubits an \emph{entangling block} consists of a layer of $k$ single-qubit quantum gates and a cyclic composition of $k$ controlled single-qubit gates, governed by an \emph{entagling range} $r$. For a given $r<k$ with $\mbox{gcd}(k,r)=1$, such cyclic composition has the form
\[
C_{r}(V_0) \, C_{2r \mbox{mod} k}(V_r) \, \cdots C_{0}(V_{(k-1)r \mbox{mod} k})
\]
\noindent where $C_c(V_t)$ denotes a controlled single-qubit gate with $t$th qubit as the target and $c$th qubit as the control.
The minimum practical number of entangling blocks was found to be $2$ (cf. also Fig.~\ref{fig:polar:code} ).

The baseline parameterization of a quantum circuit assumed that all the parameters (rotation angles) occurring in individual gates were independent. However, we also experimented with the \emph{parameter tying} strategies where there have been only $4\,k$ independent parameters shared across all the entangling blocks.

\subsubsection{Hyperparameter sweeps: quantum}

Candidate quantum circuits and feasible training options form a vast search space.
An individual quantum circuit is defined by the following hyperparameters: (1) The number of qubits $k$ (we used $k \in 2..7$ for qubit encoding and $k \in \{2,3\}$ for amplitude encoding); (2) The number of entangling blocks (in $2..5$ in most of experiments); (3) parameter tying switch ($\mbox{true}$/$\mbox{false}$).

On top of a choice of a quantum circuit, an individual training/prediction experiment required the following choices:
(1)  The width $L$ of the long memory window (discussed separately below);
(2)  The number of parameter restarts (parameters \emph{seeds});
(3) Approximation tolerances;
(4)  A cap on the number of training epochs (resp. on the number of parameter passes for the coordinate ascent method);
(5)  Learning rate (stochastic gradient descent only);
(6)  Minibatch size (stochastic gradient).

The methods were all implemented in the quantum programming language Q\# and experiments were carried out using the Microsoft Quantum Development Kit and the full-state quantum simulator it exposes \cite{QSharp2019}.

In order to cover a reasonable subset of the hyperparameter search space we assembled individual training/validation/prediction instances into large pools of asynchronous tasks deployed onto a cluster with $1000$ cores.  Traditional postprocessing was used to collect the prediction statistics.

\subsubsection{Hyperparameter sweeps: classical}

The top level variability in classical models for the bitstream prediction was around the choice of a core machine learning method. We evaluated five traditional methods, namely:
logistic regression, Naive Bayes, nearest neighbors, random forest, and Support Vector Machines. We have also evaluated six different Neural Network geometries.
While more traditional off-the-shelf methods have been used with default hyperparameter settings, the Neural Networks have been run with variability in (1) Learning rates, (2) Minibatch sizes, and (3) Activation methods.

The training window (long memory) width $L$ and $d$-gram depth (short memory depth) $d$ have been the two common hyperparameters for all the classical models. In all cases the inputs have been perceived
as $d$-dimensional feature vectors. In that sense the input representations have been a moral equivalent of the amplitude encoding in quantum-assisted analysis.

The instances using the scikit-learn tools had been pooled as asynchronous tasks and deployed to a cluster with $1000$ cores.
The \emph{Mathematica}-based instances have been executed on a $20$-core desktop with $20$-thread parallelization.

Multimodel predictors had been simulated during the classical postprocessing by either (a) model selection based on validation scores or, (b) simulated model boosting.

\section{Simulation Results}\label{subsec:sim:results}

In our experimentation, for a selected value of $L$ we extracted approximately $1000$ contiguous bitstream segments of length $L+5$ from human-generated bitstreams for each candidate $L$.
Denote these sets of segments as $\mbox{segments}(L)$ for convenience.

Each particular \emph{simulation experiment} was defined by a complete characterization of a predictor model (including values of all the hyperparameters) and the width $L$ of the intended training window. For each segment $s \in \mbox{segments}(L)$, the corresponding model was trained using the first $L$ bits of the segment $s$ and scored on the last 5 bits of the segment. In the predictor setups, where model selection was required, the selection was performed to maximize the training score.

The \emph{accuracy score} of an (experiment, segment) pair $(P,s)$  was given by $score(P,s)=\#correct/5$ where the $\#correct$  is the number of held out bits of the segment correctly predicted in that experiment.

Accuracy score for a particular predictor P given the training window width $L$ was characterized by the $\mu(P,L)=$ mean of $score(P,s)$ over the ensemble $\mbox{segments}(L)$  and by the $\sigma(P,L)=$ standard deviation of $score(P,s)$ over that ensemble.

Based on the intuition developed in subsection \ref{sub:sub:collision:statistics} $\mu_{tt}=0.62$ is a reasonable target threshold. As we will see below, there is a smaller but robust subset of predictor types (both classical and quantum) that are somewhat likely or highly likely meet or exceed this threshold.

For the sake of readability, out of the massive set of simulation results collected over extended matrix of model types and setting, we retain for the discussion only such that are comparable with the $\mu_{tt}=0.62$ mean accuracy target. These model types and setting are discussed in the following subsections.

 Emulation results are assembled in small tables, where the rows correspond to different values of the $d$-gram depth and the columns correspond to the different values of the training window width $L$. It should be noted that models with $L < 100$ were seen to underperform that accuracy threshold and models using $d>4$ have been outperformed with models that had $d \in \{3,4\}$. This demonstrates that the humans' attention window in our data collection experiment was shorter than we would have initially guessed.

In order to provide a broader context, we cite experimental accuracy metrics for selected underperforming predictors in the Appendix \ref{appendix:underperf}.

It is also notable that experiments with model boosting vs. model selection did not produce any statistically significant differentiation between the two prediction accuracy statistics. Therefore the reported results below pertain to the pure unboosted models only.

\subsection{Traditional predictors}

We experimented with the full stack of Machine Learning (ML) tools from the Python \emph{scikit-learn} and \emph{Mathematica} edition 12.
Eventually, only Logistic Regression (LR) and Neural Networks (NN) we able to achieve the competitive prediction accuracy threshold of $\mu=0.62$. LR appeared to have been somewhat more robust and accurate in \emph{Mathematica} and NN solutions - in \emph{scikit-learn}. Tables below summarize the estimated means and standard deviations for the accuracy given selected $(d,L)$ pairs.

\begin{table}[h!] \caption{Mean and standard deviation $(\mu, \sigma)$ of test bit-prediction accuracy of logistic regression, based on the \emph{Mathematica} $\mbox{Classify[*,Rule[Method,"LogisticRegression"]]}$, for $d$-grams (history) of lengths $3$, $4$, and $7$.}
\medskip
\begin{tabular}{cccc}
\hline\hline\\[-1ex]
LR & $L=100$ & $L=125$ & $L=150$ \\[0.5ex]
\hline\\[-0.5ex]
$d=3$ & (0.63,0.232) & (0.634,0.232) & \bf{(0.637,0.228)} \\
$d=4$ & (0.619,0.24) & (0.629,0.234) & (0.627, 0.232) \\
$d=7$ & (0.61,0.24) & (0.621,0.232) & (0.624, 0.228) \\ [1ex]
\hline
\hline
\end{tabular}
\label{table:logistic:regression}
\end{table}

 We evaluated an extended array of NN geometries out of which the geometries with two small hidden layers and the best-performing "Softmax" activation.

\begin{table}[h!] \caption{Mean and standard deviation $(\mu, \sigma)$ of test bit-prediction accuracy for various \emph{scikit-learn} Neural Networks with two hidden layers, for $d$-grams (history) of lengths $3$, $4$, and $7$.}
\medskip
\begin{tabular}{cccc}
\hline\hline\\[-1ex]
NN & $L=100$ & $L=125$ & $L=150$ \\[0.5ex]
\hline\\[-0.5ex]
$d=3$ & (0.629,0.23) & \bf{(0.638,0.22)} & (0.632,0.22) \\
$d=4$ & (0.595,0.234) & (0.619,0.235) & (0.6,0.225) \\
$d=7$ & (0.521,0.243) & (0.551,0.245) & (0.53,0.244) \\ [1ex]
\hline
\hline
\end{tabular}
\label{table:scikit:NN}
\end{table}

It is clear from the the bottom row of the table that NN classifiers tend to significantly overfit when the 7-grams are used, while being perfectly competitive on 3-grams.

\subsection{Quantum-assisted classifiers} \label{subsub:quantum:assisted:classifiers}

Here we report results for only two quantum-assisted classifier circuit geometries, both using the \emph{amplitude encoding} of bits streams. Exhaustive experiments with quantum circuit-centric classifiers based on qubit encoding did not furnish solutions capable of consistently meeting the target prediction accuracy threshold $\mu_{tt}=0.62$. With the use of the amplitude encoding we only needed two qubits to encode the 3-grams and only three qubits to encode 7-grams.

The 2-qubit circuit however was trimmed to 8 parameters to avoid overfitting, and represented as
\begin{eqnarray*}\label{eq:2:qubit:circuit}
M_1 \, I\otimes (R_X(\theta_8) R_Z(\theta_7)) \, C_{01}(R_X(\theta_6)) \, C_{10}(R_X(\theta_5)) \\
(R_X(\theta_3) \otimes R_X(\theta_4)) \, (R_Z(\theta_1) \otimes R_Z(\theta_2))
\end{eqnarray*}
\noindent where $R_X$, $R_Z$ are rotation around $X$ and $Z$ respectively.

\begin{table}[h!] \caption{Mean and standard deviation of test bit-prediction accuracy for circuit-centric quantum classifiers with the qubit counts ($q$) $2$ and $3$. The 2-qubit scheme naturally encodes $3$-grams. For $q=3$ we used encoding with $d=7$. }
\medskip
\begin{tabular}{cccc}
\hline\hline\\[-1ex]
QC & $L=100$ & $L=125$ & $L=150$ \\[0.5ex]
\hline\\[-1ex]
$q=2$ & (0.623,0.231) & \bf{(0.639,0.22)} & (0.624,0.23) \\
$q=3$ & (0.618, 0.231) & (0.623,0.235) & (0.619, 0.236) \\ [1ex]
\hline\hline
\end{tabular}
\label{table:quantum:circuit}
\end{table}

The estimates for the mean and standard deviation of the bit-prediction accuracy are presented in Table \ref{table:quantum:circuit}.

\subsection{Comparative overview}

As per Tables~\ref{table:logistic:regression}--\ref{table:quantum:circuit}, our experiments deliver accuracy estimates with standard deviations in the $(0.22,0.24)$ range over an ensemble of 1000 experiments. The best results in the tables are seen to improve on the $\mu_{tt}=0.62$ threshold with high confidence. (Bests results - with confidence score in the $[0.99,0.997]$ range assuming normality.)

Since $d$-gramm depth of 3 appears to be the most robust for all models, table \ref{table:d:3} below compares per-method accuracy statistics for all the predictors at $d=3$

\begin{table}[h!] \caption{Comparison of mean and standard deviation $(\mu, \sigma)$ of test bit-prediction accuracy for the best bit-predictors for two classical methods (LR, NN) and one quantum method (QC). LR stands for Logistic Regression with $d=3$, NN stands for Neural Networks with $d=3$, QC stands for quantum circuit-centric classifiers with $d=2$, and
 $L$ is the training window width.}
\medskip
\begin{tabular}{cccc}
\hline\hline\\[-1ex]
$d=3$ & $L=100$ & $L=125$ & $L=150$  \\ [0.5ex]
\hline\\[-1ex]
LR & (0.63,0.232) & (0.634,0.232) & \bf{(0.637,0.228)} \\
NN & (0.629,0.23) & \bf{(0.638,0.22)} & (0.632,0.22) \\
QC & (0.623,0.231) & \bf{(0.639,0.22)} & (0.624,0.23) \\ [1ex]
\hline
\hline
\end{tabular}
\label{table:d:3}
\end{table}

Unfortunately, due to relatively large variances it is impossible to statistically differentiate between various predictors rated in the above tables with sufficient confidence.

\section{Conclusion}
We completed a comparative study of classical versus quantum predictors that drive computer simulation of human-generated bitstreams.
The bitstreams used in the study have been generated under the ``randomization'' imperative that by design made accurate prediction hard.

The presented statistical data is based on forecasting bits in bitstreams of length 1000, collected from a group of $34$ volunteers. Our findings seem to indicate that, on average, the next bit can be accurately forecast in about $64$ percent of cases by use of trained quantum circuits that perform the prediction.

Our initial hypothesis have been that the use of quantum correlations for predicting human choices gives a distinct predictive advantage over the use of only classical correlations. However, this hypothesis could not be ascertained or rejected in the context of the present study. It appears that the conditional distribution of the follow on bit in the context can be just as accurately described by classical predictors such as logistic autoregression or simple neural network. There are possible principled as well as technical explanations for this outcome, which will be the topic of future research.

\vspace{\baselineskip}
\section*{Acknowledgements}
The authors thank Guang Hao Low for discussions and help with deploying experimental simulations in Azure. AB also wishes to thank Etienne Bernard, Daniel Lichtblau and Jerome Louradour for a crash intro into \emph{Mathematica}'s machine learning tools.

\bibliography{BitStreamPredictione}

\begin{thebibliography}{16}%
\makeatletter
\providecommand \@ifxundefined [1]{%
 \@ifx{#1\undefined}
}%
\providecommand \@ifnum [1]{%
 \ifnum #1\expandafter \@firstoftwo
 \else \expandafter \@secondoftwo
 \fi
}%
\providecommand \@ifx [1]{%
 \ifx #1\expandafter \@firstoftwo
 \else \expandafter \@secondoftwo
 \fi
}%
\providecommand \natexlab [1]{#1}%
\providecommand \enquote  [1]{``#1''}%
\providecommand \bibnamefont  [1]{#1}%
\providecommand \bibfnamefont [1]{#1}%
\providecommand \citenamefont [1]{#1}%
\providecommand \href@noop [0]{\@secondoftwo}%
\providecommand \href [0]{\begingroup \@sanitize@url \@href}%
\providecommand \@href[1]{\@@startlink{#1}\@@href}%
\providecommand \@@href[1]{\endgroup#1\@@endlink}%
\providecommand \@sanitize@url [0]{\catcode `\\12\catcode `\$12\catcode
  `\&12\catcode `\#12\catcode `\^12\catcode `\_12\catcode `\%12\relax}%
\providecommand \@@startlink[1]{}%
\providecommand \@@endlink[0]{}%
\providecommand \url  [0]{\begingroup\@sanitize@url \@url }%
\providecommand \@url [1]{\endgroup\@href {#1}{\urlprefix }}%
\providecommand \urlprefix  [0]{URL }%
\providecommand \Eprint [0]{\href }%
\providecommand \doibase [0]{http://dx.doi.org/}%
\providecommand \selectlanguage [0]{\@gobble}%
\providecommand \bibinfo  [0]{\@secondoftwo}%
\providecommand \bibfield  [0]{\@secondoftwo}%
\providecommand \translation [1]{[#1]}%
\providecommand \BibitemOpen [0]{}%
\providecommand \bibitemStop [0]{}%
\providecommand \bibitemNoStop [0]{.\EOS\space}%
\providecommand \EOS [0]{\spacefactor3000\relax}%
\providecommand \BibitemShut  [1]{\csname bibitem#1\endcsname}%
\let\auto@bib@innerbib\@empty
\bibitem [{\citenamefont {Fisher}(2015)}]{fisher15}%
  \BibitemOpen
  \bibfield  {author} {\bibinfo {author} {\bibfnamefont {M.~P.~A.}\
  \bibnamefont {Fisher}},\ }\href@noop {} {\bibfield  {journal} {\bibinfo
  {journal} {Annals of Physics}\ }\textbf {\bibinfo {volume} {362}},\ \bibinfo
  {pages} {593} (\bibinfo {year} {2015})}\BibitemShut {NoStop}%
\bibitem [{\citenamefont {Aerts}\ \emph {et~al.}(2015)\citenamefont {Aerts},
  \citenamefont {Sozzo},\ and\ \citenamefont {Veloz}}]{aerts15}%
  \BibitemOpen
  \bibfield  {author} {\bibinfo {author} {\bibfnamefont {D.}~\bibnamefont
  {Aerts}}, \bibinfo {author} {\bibfnamefont {S.}~\bibnamefont {Sozzo}}, \ and\
  \bibinfo {author} {\bibfnamefont {T.}~\bibnamefont {Veloz}},\ }\href@noop {}
  {\bibfield  {journal} {\bibinfo  {journal} {International Journal of
  Theoretical Physicss}\ }\textbf {\bibinfo {volume} {54}},\ \bibinfo {pages}
  {4557} (\bibinfo {year} {2015})}\BibitemShut {NoStop}%
\bibitem [{\citenamefont {Halpern}\ and\ \citenamefont
  {Crosson}(2019)}]{halpern17}%
  \BibitemOpen
  \bibfield  {author} {\bibinfo {author} {\bibfnamefont {N.~Y.}\ \bibnamefont
  {Halpern}}\ and\ \bibinfo {author} {\bibfnamefont {E.}~\bibnamefont
  {Crosson}},\ }\href@noop {} {\bibfield  {journal} {\bibinfo  {journal}
  {Annals of Physics}\ }\textbf {\bibinfo {volume} {407}},\ \bibinfo {pages}
  {92} (\bibinfo {year} {2019})}\BibitemShut {NoStop}%
\bibitem [{\citenamefont {Merrill}(2018)}]{merril2018}%
  \BibitemOpen
  \bibfield  {author} {\bibinfo {author} {\bibfnamefont {N.}~\bibnamefont
  {Merrill}},\ }\href@noop {} {\emph {\bibinfo {title} {{'Aaronson oracle'
  project}}}},\ \bibinfo {type} {Tech. Rep.}\ (\bibinfo  {institution} {UC
  Berkeley},\ \bibinfo {address} {https://github.com/elsehow/aaronson-oracle},\
  \bibinfo {year} {2018})\BibitemShut {NoStop}%
\bibitem [{\citenamefont {Schuld}\ \emph {et~al.}(2018)\citenamefont {Schuld},
  \citenamefont {Bocharov}, \citenamefont {Svore},\ and\ \citenamefont
  {Wiebe}}]{schuld18}%
  \BibitemOpen
  \bibfield  {author} {\bibinfo {author} {\bibfnamefont {M.}~\bibnamefont
  {Schuld}}, \bibinfo {author} {\bibfnamefont {A.}~\bibnamefont {Bocharov}},
  \bibinfo {author} {\bibfnamefont {K.~M.}\ \bibnamefont {Svore}}, \ and\
  \bibinfo {author} {\bibfnamefont {N.}~\bibnamefont {Wiebe}},\ }\href@noop {}
  {\bibfield  {journal} {\bibinfo  {journal} {arXiv preprint arXiv:1804.00633}\
  } (\bibinfo {year} {2018})}\BibitemShut {NoStop}%
\bibitem [{\citenamefont {Bottou}(2004)}]{bottou2004}%
  \BibitemOpen
  \bibfield  {author} {\bibinfo {author} {\bibfnamefont {L.}~\bibnamefont
  {Bottou}},\ }\href@noop {} {\emph {\bibinfo {title} {Stochastic Learning}}},\
  \bibinfo {edition} {advanced lectures on machine learning}\ ed.,\ Vol.\
  \bibinfo {volume} {3176}\ (\bibinfo  {publisher} {Springer, LNAI},\ \bibinfo
  {year} {2004})\BibitemShut {NoStop}%
\bibitem [{\citenamefont {Ostaszewski}\ \emph {et~al.}(2019)\citenamefont
  {Ostaszewski}, \citenamefont {Grant},\ and\ \citenamefont
  {Benedetti}}]{OstaszewskiEtAl}%
  \BibitemOpen
  \bibfield  {author} {\bibinfo {author} {\bibfnamefont {M.}~\bibnamefont
  {Ostaszewski}}, \bibinfo {author} {\bibfnamefont {E.}~\bibnamefont {Grant}},
  \ and\ \bibinfo {author} {\bibfnamefont {M.}~\bibnamefont {Benedetti}},\
  }\href@noop {} {\bibfield  {journal} {\bibinfo  {journal} {arXiv preprint
  arXiv:11905.09692}\ } (\bibinfo {year} {2019})}\BibitemShut {NoStop}%
\bibitem [{\citenamefont {Bocharov}\ \emph {et~al.}(2020)\citenamefont
  {Bocharov}, \citenamefont {Roetteler},\ and\ \citenamefont
  {Svore}}]{toappear20}%
  \BibitemOpen
  \bibfield  {author} {\bibinfo {author} {\bibfnamefont {A.}~\bibnamefont
  {Bocharov}}, \bibinfo {author} {\bibfnamefont {M.}~\bibnamefont {Roetteler}},
  \ and\ \bibinfo {author} {\bibfnamefont {K.~M.}\ \bibnamefont {Svore}},\
  }\href@noop {} {\bibfield  {journal} {\bibinfo  {journal} {(Manuscript)}\ }
  (\bibinfo {year} {2020})}\BibitemShut {NoStop}%
\bibitem [{\citenamefont {Levine}\ \emph {et~al.}(2019)\citenamefont {Levine},
  \citenamefont {Sharir}, \citenamefont {Cohen},\ and\ \citenamefont
  {Shashua}}]{fourAuthors19}%
  \BibitemOpen
  \bibfield  {author} {\bibinfo {author} {\bibfnamefont {Y.}~\bibnamefont
  {Levine}}, \bibinfo {author} {\bibfnamefont {O.}~\bibnamefont {Sharir}},
  \bibinfo {author} {\bibfnamefont {N.}~\bibnamefont {Cohen}}, \ and\ \bibinfo
  {author} {\bibfnamefont {A.}~\bibnamefont {Shashua}},\ }\href@noop {}
  {\bibfield  {journal} {\bibinfo  {journal} {Phys. Rev. Lett.}\ }\textbf
  {\bibinfo {volume} {122}},\ \bibinfo {pages} {065301} (\bibinfo {year}
  {2019})}\BibitemShut {NoStop}%
\bibitem [{\citenamefont {Deng}\ \emph {et~al.}(2017)\citenamefont {Deng},
  \citenamefont {Ki},\ and\ \citenamefont {Das~Sarma}}]{deng17}%
  \BibitemOpen
  \bibfield  {author} {\bibinfo {author} {\bibfnamefont {D.-L.}\ \bibnamefont
  {Deng}}, \bibinfo {author} {\bibfnamefont {X.}~\bibnamefont {Ki}}, \ and\
  \bibinfo {author} {\bibfnamefont {S.}~\bibnamefont {Das~Sarma}},\ }\href@noop
  {} {\bibfield  {journal} {\bibinfo  {journal} {Physical Review X.}\ }\textbf
  {\bibinfo {volume} {7}},\ \bibinfo {pages} {021021} (\bibinfo {year}
  {2017})}\BibitemShut {NoStop}%
\bibitem [{\citenamefont {Aharonov}\ \emph {et~al.}(2006)\citenamefont
  {Aharonov}, \citenamefont {Jones},\ and\ \citenamefont {Landau}}]{AJL2005}%
  \BibitemOpen
  \bibfield  {author} {\bibinfo {author} {\bibfnamefont {D.}~\bibnamefont
  {Aharonov}}, \bibinfo {author} {\bibfnamefont {V.}~\bibnamefont {Jones}}, \
  and\ \bibinfo {author} {\bibfnamefont {Z.}~\bibnamefont {Landau}},\
  }\href@noop {} {\bibfield  {journal} {\bibinfo  {journal} {STOC 2006}\ }
  (\bibinfo {year} {2006})}\BibitemShut {NoStop}%
\bibitem [{QSh(2019)}]{QSharp2019}%
  \BibitemOpen
  \href@noop {} {\emph {\bibinfo {title} {{Quantum basics with Q\#}}}},\
  \bibinfo {type} {Tech. Rep.}\ (\bibinfo  {institution} {Microsoft Quantum
  Systems},\ \bibinfo {address}
  {https://docs.microsoft.com/en-us/quantum/quickstart},\ \bibinfo {year}
  {2019})\BibitemShut {NoStop}%
\bibitem [{\citenamefont {Figurska}\ \emph {et~al.}(2008)\citenamefont
  {Figurska}, \citenamefont {Stanczyk},\ and\ \citenamefont
  {Kulesza}}]{rng2008}%
  \BibitemOpen
  \bibfield  {author} {\bibinfo {author} {\bibfnamefont {M.}~\bibnamefont
  {Figurska}}, \bibinfo {author} {\bibfnamefont {M.}~\bibnamefont {Stanczyk}},
  \ and\ \bibinfo {author} {\bibfnamefont {K.}~\bibnamefont {Kulesza}},\
  }\href@noop {} {\bibfield  {journal} {\bibinfo  {journal} {Med Hypotheses}\
  }\textbf {\bibinfo {volume} {70(1)}},\ \bibinfo {pages} {182 } (\bibinfo
  {year} {2008})}\BibitemShut {NoStop}%
\bibitem [{\citenamefont {Jokar}\ and\ \citenamefont
  {Mikaili}(2012)}]{rng2012}%
  \BibitemOpen
  \bibfield  {author} {\bibinfo {author} {\bibfnamefont {E.}~\bibnamefont
  {Jokar}}\ and\ \bibinfo {author} {\bibfnamefont {M.}~\bibnamefont
  {Mikaili}},\ }\href@noop {} {\bibfield  {journal} {\bibinfo  {journal} {J Med
  Signals Sens}\ }\textbf {\bibinfo {volume} {2(2)}},\ \bibinfo {pages} {82 }
  (\bibinfo {year} {2012})}\BibitemShut {NoStop}%
\bibitem [{sci(2019)}]{scikit2019}%
  \BibitemOpen
  \href@noop {} {\emph {\bibinfo {title} {Scikit Learn version 0.22.1
  documentation}}},\ \bibinfo {type} {Tech. Rep.}\ (\bibinfo  {institution}
  {https://scikit-learn.org/stable/modules/},\ \bibinfo {address}
  {neural\_networks\_supervised.html\#classification},\ \bibinfo {year}
  {2019})\BibitemShut {NoStop}%
\bibitem [{mma(2019)}]{mmaClassify2019}%
  \BibitemOpen
  \href@noop {} {\emph {\bibinfo {title} {{Wolfram Mathematica 12
  Documentation}}}},\ \bibinfo {type} {Tech. Rep.}\ (\bibinfo  {institution}
  {Wolfram Research},\ \bibinfo {address}
  {https://reference.wolfram.com/language/ref/Classify.html},\ \bibinfo {year}
  {2019})\BibitemShut {NoStop}%
\end{thebibliography}%

\appendix
\section{Accuracy metrics for selected underperforming predictors} \label{appendix:underperf}

As we have stated in the main body of text, the majority of classical predictive models we have been evaluating, significantly underperformed the target mean accuracy threshold of \textbf{0.62}.
In order to illustrate typical underperforming behaviors we present the bit prediction accuracy statistics for a selection of such predictive models. The multitude of models we have been evaluating with varying degree of success give some empirical certainty that said accuracy threshold is dictated by statistical properties of the data collection. The threshold appears to be hard to improve on with either traditional or non-traditional predictive strategies (such as variational quantum circuits).

\subsection{Accuracy metrics for $d$-gram oracle predictor}

Here we report the mean prediction accuracies for the $d$-gram oracle predictor for sufficient matrix of $d$ and $L$ (the training window width). Although the outcomes for various choices of ($d$,$L$) cannot be differentiated with sufficient statistical significance, it is somewhat likely that the oracle method favors the $d$-gram depth of $d=4$ . Overall the method significantly underperforms the target accuracy threshold $\mu_{tt}=0.62$.

\begin{table}[h!]
\caption{Prediction accuracies for $d$-gram oracle predictor. $L$ is the width of the long memory window over which the conditional distributions of the follow on bit have been collected.}
\label{table:d:gram:oracle}
\vskip 0.15in
\begin{center}
\begin{small}
\begin{sc}
\begin{tabular}{lcccr}
\toprule
$(\mu,\sigma)$ & $L=100$  & $L=125$  & $L=150$ \\
\midrule
$d=3$    &  (0.578,0.22) & (0.604.0.229) & (0.572,0.218)\\
$d=4$ &  (0.579,0.226)& $(0.602,0.227)$ & (0.579,0.218)\\
$d=7$    & (0.578,0.222)& (0.58,0.22) & (0.579,0.218)\\
\bottomrule
\end{tabular}
\end{sc}
\end{small}
\end{center}
\vskip -0.1in
\end{table}

\subsection{Accuracy metrics for Support Vector Machine classifiers}

Table \ref{table:SVM} below summarizes the accuracies for the prediction of follow on bit using $\mbox{Classify[*,"SupportVectorMachine"]}$ function of \emph{Mathematica} 12. For the short memory depth $d$ the corresponding $d$-grams were treated as data vectors for the SVM method.

\begin{table}[h!]
\caption{Prediction accuracies for Support Vector Machine classifiers. $L$ stands for the width of the training ('long memory') window.}
\label{table:SVM}
\vskip 0.15in
\begin{center}
\begin{small}
\begin{sc}
\begin{tabular}{lcccr}
\toprule
$(\mu,\sigma)$ & $L=100$  & $L=125$  & $L=150$ \\
\midrule
$d=3$    &  (0.571,0.244) & (0.583,0.239) & (0.57,0.246)\\
$d=4$ &  (0.574,0.246)& $(0.579,0.247)$ & (0.592,0.242)\\
$d=7$    & (0.563,0.246)& (0.574,0.242) & (0.584,0.239)\\
\bottomrule
\end{tabular}
\end{sc}
\end{small}
\end{center}
\vskip -0.1in
\end{table}

\subsection{Single layer classifiers with hidden layer of size $d$.}

The tables \ref{table:d:NN:Logit},\ref{table:d:NN:Tanh},\ref{table:d:NN:ReLU} present the prediction accuracy statistics for single layer classifiers with one hidden layer of sizes $d$. The accuracies appear to be significantly lower than those achieved by 2-layer classifiers, as summarized in the main text and significantly lower than the target accuracy threshold of $0.62$. The tables below present results for three different choices of the nonlinear activation function. The statistics is collected using 3-layer neural network classifiers built with \emph{Mathematica} 12 machine learning tools. In a majority of the $(d,L)$ configurations the \emph{scikit-learn} multilayer classifiers have been also evaluated leading to essentially similar or visually inferior results.

\begin{table}[h!]
\caption{Prediction accuracies for single layer NN classifier with $\mbox{Logit}$ activation.}
\label{table:d:NN:Logit}
\vskip 0.15in
\begin{center}
\begin{small}
\begin{sc}
\begin{tabular}{lcccr}
\toprule
$(\mu,\sigma)$ & $L=100$  & $L=125$  & $L=150$ \\
\midrule
$d=3$    &  (0.57,0.237) & (0.569, 0.243) & (0.572,0.242)\\
$d=4$ &  (0.571,0.24)& $(0.583,0.231)$ & (0.585,0.234)\\
$d=7$    & (0.575,0.232)& (0.584,0.233) & (0.584,0.23)\\
\bottomrule
\end{tabular}
\end{sc}
\end{small}
\end{center}
\vskip -0.1in
\end{table}

\begin{table}[h!]
\caption{Prediction accuracies for single layer NN classifier with $Tanh$ activation.}
\label{table:d:NN:Tanh}
\vskip 0.15in
\begin{center}
\begin{small}
\begin{sc}
\begin{tabular}{lcccr}
\toprule
$(\mu,\sigma)$ & $L=100$  & $L=125$  & $L=150$ \\
\midrule
$d=3$    &  (0.582,0.239) & (0.593.0.232) & (0.591,0.23)\\
$d=4$ &  (0.585,0.24)& $(0.602,0.23)$ & (0.607,0.235)\\
$d=7$    & (0.579,0.235)& (0.591,0.235) & (0.591,0.23)\\
\bottomrule
\end{tabular}
\end{sc}
\end{small}
\end{center}
\vskip -0.1in
\end{table}

\begin{table}[h!]
\caption{Prediction accuracies for single layer NN classifier with $SELU$ activation.}
\label{table:d:NN:ReLU}
\vskip 0.15in
\begin{center}
\begin{small}
\begin{sc}
\begin{tabular}{lcccr}
\toprule
$(\mu,\sigma)$ & $L=100$  & $L=125$  & $L=150$ \\
\midrule
$d=3$    &  (0.593,0.237) & (0.599,.0.228) & (0.595,0.23)\\
$d=4$ &  (0.578,0.246)& $(0.583,0.239)$ & (0.595,0.24)\\
$d=7$    & (0.574,0.233)& (0.598,0.241) & (0.597,0.24)\\
\bottomrule
\end{tabular}
\end{sc}
\end{small}
\end{center}
\vskip -0.1in
\end{table}

\subsection{Multilayer classifiers with three hidden layers.}

The tables below present the prediction accuracy statistics for 3-layer classifiers with layer sizes $(d,d,d)$. The accuracies appear to be significantly lower than those achieved by 2-layer classifiers, as summarized in the main text and significantly lower than the target accuracy threshold of $0.62$. The tables below present results for three different choices of the nonlinear activation function. The statistics is collected using 3-layer neural network classifiers built with \emph{Mathematica} 12 machine learning tools. In a majority of the $(d,L)$ configurations the \emph{scikit-learn} multilayer classifiers have been also evaluated leading to essentially similar or visually inferior results.

\begin{table}[h!]
\caption{Prediction accuracies for 3-layer NN classifier with $\mbox{Logit}$ activation.}
\label{table:ddd:NN:Logit}
\vskip 0.15in
\begin{center}
\begin{small}
\begin{sc}
\begin{tabular}{lcccr}
\toprule
$(\mu,\sigma)$ & $L=100$  & $L=125$  & $L=150$ \\
\midrule
$d=3$    &  (0.533,0.24) & (0.536.0.242) & (0.541,0.241)\\
$d=4$ &  (0.561,0.234)& $(0.572,0.236)$ & (0.584,0.234)\\
$d=7$    & (0.57,0.237)& (0.578,0.244) & (0.573,0.236)\\
\bottomrule
\end{tabular}
\end{sc}
\end{small}
\end{center}
\vskip -0.1in
\end{table}

\begin{table}[h!]
\caption{Prediction accuracies for 3-layer NN classifier with $Tanh$ activation.}
\label{table:ddd:NN:Tanh}
\vskip 0.15in
\begin{center}
\begin{small}
\begin{sc}
\begin{tabular}{lcccr}
\toprule
$(\mu,\sigma)$ & $L=100$  & $L=125$  & $L=150$ \\
\midrule
$d=3$    &  (0.598,0.232) & (0.596.0.233) & (0.591,0.235)\\
$d=4$ &  (0.607,0.226)& $(0.609,0.224)$ & (0.617,0.227)\\
$d=7$    & (0.586,0.238)& (0.574,0.238) & (0.598,0.239)\\
\bottomrule
\end{tabular}
\end{sc}
\end{small}
\end{center}
\vskip -0.1in
\end{table}

\begin{table}[h!]
\caption{Prediction accuracies for 3-layer NN classifier with $SELU$ activation.}
\label{table:ddd:NN:ReLU}
\vskip 0.15in
\begin{center}
\begin{small}
\begin{sc}
\begin{tabular}{lcccr}
\toprule
$(\mu,\sigma)$ & $L=100$  & $L=125$  & $L=150$ \\
\midrule
$d=3$    &  (0.594,0.23) & (0.592,.0.236) & (0.594,0.233)\\
$d=4$ &  (0.586,0.24)& $(0.582,0.238)$ & (0.594,0.238)\\
$d=7$    & (0.588,0.242)& (0.591,0.24) & (0.603,0.232)\\
\bottomrule
\end{tabular}
\end{sc}
\end{small}
\end{center}
\vskip -0.1in
\end{table}

\end{document}